\documentclass[acmtog]{acmart}

\AtBeginDocument{%
  \providecommand\BibTeX{{%
    \normalfont B\kern-0.5em{\scshape i\kern-0.25em b}\kern-0.8em\TeX}}}

\DeclareMathOperator{\ft}{FT}
\DeclareMathOperator{\ift}{FT^{-1}}

\usepackage{float}

\begin{document}

\title{Global displacement induced by rigid motion simulation during MRI acquisition.}

\author{Ghiles Reguig}

\email{ghiles.reguig@gmail.com}
\affiliation{%
  \institution{Paris Brain Institute – ICM, Siemens Healthineers.}
  \city{Paris}
  \country{France}
}

\author{Marc Lapert}
\affiliation{
  \institution{Siemens Healthinners}
  \city{Paris}
  \country{France}}

\author{Stéphane Lehéricy}
\affiliation{%
  \institution{Paris Brain Institute – ICM, INSERM U 1127, CNRS UMR 7225, Sorbonne Université; Department of Neuroradiology, Pitié-Salpêtrière Hospital, Public Assistance - Paris Hospitals (AP-HP), Paris, France}
  \city{Paris}
  \country{France}
}

\author{Romain Valabregue}
\affiliation{%
  \institution{Paris Brain Institute – ICM, INSERM U 1127, CNRS UMR 7225, Sorbonne Université}
  \city{Paris}
  \country{France}
}
\email{romain.valabregue@upmc.fr}

\renewcommand{\shortauthors}{G. Reguig, et al.}

\newpage

\begin{abstract}
    \section*{\textbf{Abstract}}

\textbf{Purpose}:  Simulation of motion in the frequency domain during the MRI acquisition process is an important tool to study the effect of object motion. It is natural in this context to rely on a voxel-to-voxel difference metric between the original image and its motion-simulated version to quantify the alteration. These metrics are very sensitive to any mis-coregistration between the two images. As we observed global displacements of the corrupted image for specific simulated motions, we focused this work on the critical choice of a reference position of the object for the motion simulation that avoids any global displacement.\\
\textbf{Methods}: We used the motion simulation framework and we studied the different proposed solutions. Classically, the position around the k-space center is considered as a reference with two analytical solutions to correct for global image shifts: demeaning the image in the Fourier domain either using the exact position at k-space center or using a weighted average around the center.\\
\textbf{Results}: We demonstrated with motion simulated examples that these analytical solutions did not yield the same results as a direct co-registration method. Indeed, short transitions around the k-space center did not induce any global displacement, in contrast with what both analytical solutions assumed. We also observed a dependence of the global shift to both the motion amplitude and the translation direction.\\
\textbf{Conclusion}: In the absence of a theoretical solution, we concluded that an extra co-registration step after motion simulation was needed in order to compare the simulation with its reference image.\\
\end{abstract}

\maketitle

\section{\textbf{Introduction}}
\label{intro}
The rigid movement of the head is the predominant artifact of 3D cerebral MRI acquisitions, due to its long acquisition time. As motion occurs while sampling the frequency domain, the effects on the image are complex and hard to predict. Quality control studies, which aimed to automatically detect motion-artifacted images (among other artifacts) always relied on a ground truth defined by an expert (\cite{alfaro2018image, esteban2017mriqc}). This restricts to a binary and noisy definition of the artifact severity with a poor inter-rater reproducibility (\~ 0.85 in \cite{esteban2017mriqc}). On the other hand, a realistic motion simulation framework provides a tool to better define the motion artifact severity, by relying on a voxel-to-voxel difference metric (L1, L2, NCC, MI, SSIM …) which quantifies the difference between the motion simulated image and its original version. Those difference metrics are also used as loss functions for deep learning approaches and constitute a key element of the learning strategy of motion detection/correction tasks \cite{shaw2020k, duffy2021retrospective, johnson2019conditional, liu2020motion, pawar2019suppressing, tamada2020motion}. 
\\ The objective of this work is to point out an important bias of those metrics as they are known to be very sensitive to co-registration errors. Indeed simulating a motion during the acquisition process may induce a global displacement of the image, due to the fact that subject position is defined relatively to a chosen reference. Changing the reference, which is equivalent to subtracting a constant position, would result in the same motion corrupted volume but with a global shift. Given a particular motion time course representing the different subject’s positions during data acquisition, the global mean displacement (or a reference position \footnote{Choosing the correct reference position is equivalent to predict the global displacement and choosing it as the reference. So both terms “reference position” and “global displacement” can be used interchangeably.}) has to be subtracted, so that the simulated motion corrupted volume remains at the same global position as in the original volume. 
\\ To the best of our knowledge, the global displacement problem has not been explicitly explored in the literature and is often bypassed in studies relying on motion simulation even though a voxel-to-voxel difference metric was used. Few works mentioned specific solutions. In the retromoco toolbox (\cite{gallichan2021retromoco}), the mean reference position is chosen at the k-space center. \cite{bazin2020sharpness} proposed to take as a reference the average position over eleven lines around the k-space center. \cite{shaw2020k} took a weighted average of the motion time course, with the weights defined as the absolute value of the Fourier Transform coefficients. Although the formulation is different, the latter is equivalent to the proposition of \cite{todd2015prospective}, where a partition-weighted integrated motion was used. A more radical solution is to avoid any motion around the k-space center as in \cite{duffy2021retrospective}. But the chosen 10\% portion around the k-space center is quite arbitrary and prevents the study of motion occurring near the k-space center. Finally, we only found one work that proposed to add an extra co-registration step of the motion-simulated image (\cite{johnson2019conditional}).
\\ Although the motion simulation is well known, different strategies were proposed for its implementation. \cite{liu2020motion} simulated only translation by adding a pseudo periodic phase shift perturbation to the FFT signal as proposed by \cite{tamada2020motion} for breathing artifacts simulation on liver imaging. It is however less suited for the rigid brain motion as it is important to also consider the effect of head rotation. A similar approach was used by \cite{duffy2021retrospective} by considering a Gaussian random walk perturbation of the k-space lines (translations and rotations).
\\ A second approach was used by \cite{johnson2019conditional, pawar2019suppressing, shaw2020k}. They assumed a piecewise motion, where the object takes n different positions during acquisition. The n motions are applied in the image domain and the ffts are computed for each resliced image. Then a new k-space is built by concatenating blocks corresponding to the duration of a given position and a final fft provides the motion-corrupted volume. Although this strategy is correct, the computation time needed to reslice each position and to compute the corresponding Fourier Transform makes it impractical for more than a few positions (maximal number of positions used is 5 in \cite{shaw2020k} and 32 in \cite{pawar2019suppressing}). This constrains the method to study only over simplified motion.
\\ The third approach that we followed in this work, which is equivalent in results to the second approach described above, relies on the known effect of motion on the k-space: a translation in the image domain corresponds to a multiplication of k-space data with linear phase ramps proportional to the translation amplitude, and rotation around the volume center in image domain corresponds to the same rotation around the DC component in k-space (\cite{gallichan2021retromoco, loktyushin2013blind, zahneisen2016homogeneous}). The main advantage of this approach is that we can consider as input, motion time course at any resolution.
\\ The first contribution of this work is to propose an open source implementation of this third approach in the torchio library \cite{perez2021torchio}, to help its use by the machine learning community. 
The second contribution is to demonstrate the importance of global displacement induced by motion simulation: we showed that none of the proposed solutions to correct it are in perfect agreement with the direct co-registration method (considered as ground truth). The first experiment was done in one dimension in order to show the problem in a simplified setting. In a second experiment, we explored the case of 3D brain MRI acquisition for the simplest motion: one transient position change. This allowed us to systematically explore the effects of four factors: the direction, the amplitude, the duration and the position in k-space where the motion occurs. Finally, we explored the more realistic case of random motion pattern and further demonstrated the need of an external co-registration step to avoid an over estimation of the artifact severity.

\section{\textbf{Methods}}
\label{methods}

\subsection{Motion representation}
\label{motion_representation}

\subsubsection{Motion Time course}
\label{mot_timecourse}
The motion time course is defined as the temporal positions of the object during acquisition. To describe the motion time course, we used Euler representation with 6 parameters (3 rotations and 3 translations) defined over time. To model realistic motion, we used the motion model as described in \cite{gallichan2021retromoco} composed of three components: a Perlin noise to model ‘slow’ random motions, a step displacement to model sudden motions and a transient motion to model swallowing movements. Two examples are given in Figure 5.
For Experiment 2, a two-position problem modeled by a transient step motion with varying durations and k-space locations was considered (see Figure \ref{figure2}).

\subsubsection{Motion amplitude, max displacement}
\label{max_amp}
For a given motion time course, the amplitude of the movement was computed as the maximum difference between all positions. For the translation part we computed the differences of the translation vector.  
\[Amplitude Translation (mm) = \max_{i, j>i} (|\overrightarrow{T_i} - \overrightarrow{T_j}|) \]
where $i$ and $j$ are the time step indices and $\overrightarrow{T}$ the translation vector. 
\\To accurately set a chosen maximum amplitude, the time course was divided by this maximal pairwise distance between all positions and multiplied by the desired amplitude. The same was applied to constrain the rotation amplitude (i.e. the norm of the vector $R_{x,y,z}$ differences). Note that this motion amplitude is independent of the chosen reference.

\subsubsection{Averaging affine transforms}
\label{affine_avg}
As we will see in section 2.2.2, a weighted average of the motion time course is needed for the reference position computation. The average is computed on each Euler parameters separately. Even though this is correct for translations, it is not accurate for rotations. Averaging rotations is indeed a very difficult task, with no analytical form as reviewed previously by \cite{hanson2020quaternion}. In machine vision (\cite{kavan2006dual}) or for entropy estimation (\cite{huggins2014comparing}), the advantage of using dual quaternions formalism was demonstrated to properly average rotations. Different averaging propositions were compared but no large differences were found for small rotations ($<10^{\circ}$). In this work, only the simple Euler parameters average was used.

\subsection{Motion artifact simulation}
\label{motion_artifact_simulation}

\subsubsection{Motion during k-space acquisition}
\label{motion_during_acquisition}

We followed the approach described by \cite{loktyushin2013blind}. We first applied the rotation on the kspace grid and performed a type 2 nufft (non uniform fast Fourier transform) (\cite{barnett2019parallel})\footnote{\url{https://github.com/flatironinstitute/finufft/}}. Then the linear phase shift proportional to the translation amplitude was added to the complex data before the final inverse fft. The derivation calculation is detailed in the appendix.
The algorithm requires having the motion time course defined on each kspace point. In this work, we limited ourselves to the study of a segmented 3D acquisition sequence of the MPRAGE type for which the frequency and the first phase dimension are acquired within a short time ($< 300 ms$). Hence, we supposed that no motion occurred while acquiring these two dimensions. It was thus only necessary to define the motion time course (or interpolate it) at the resolution of the third phase direction (y axis), and then to apply the motion on k-space planes instead of k-space points. 
We propose an open source implementation within the torchio library \cite{perez2021torchio}, which regroups different medical imaging augmentation tools in a pytorch compatible environment\footnote{At the time of writing, this is not yet incorporated, but a compatible version is available on our github repository : \url{https://github.com/romainVala/torchio}}.

\subsubsection{Computation of the induced global displacement}
\label{induced_global_disp}
Two different propositions have been made to define a reference position: either the position at the k-space center \cite{gallichan2021retromoco} or a weighted average based on the Fast Fourier Transform (FFT) coefficients of the image \cite{shaw2020k, todd2015prospective} from which we derived three ways of computing the reference: center, wFT\_mean and wFT2\_mean. An extra reference position (coreg\_shift), derived from a co-registration algorithm, was used as the ground truth.
\begin{description}
    \item [center] the reference was the position at the k-space center.
    
    \item [wFT\_mean (or wFT)] the reference was a weighted average of the motion time course, where the coefficients wFT were the square root of the FFT signal power of the original image at each phase step,
    \[wFT_y=\sqrt{\sum_{x,z}FT^2_{x y z}} \] 
    where $FT_{xyz}$ were the discrete Fourier Transform coefficients. All weights on the x and z dimensions were summed as the simulated motion only occurred in the slowest sampling dimension (y axis). This could be easily extended to the x, y and z dimensions if the motion was defined on all k-space points. Although different, the formulation was equivalent to the proposition made by \cite{shaw2020k}.
    
    \item [wFT2\_mean (or wFT2)] the reference was computed similarly to wFT but using squared weights. Figure 2 shows that this was very similar to the center choice.
    
    \item [coreg\_shift] the reference was the co-registration between the motion simulated image and the original one. This was used as ground truth. For the one-dimensional (1D) experiment the shift was measured by minimizing the L1 difference for all possible pixel shifts. In the 3D experiment, the rigid co-registration from Elastix software (\cite{klein2009elastix}) was used.
\end{description}

As stated in the introduction, this global displacement needs to be considered only for a voxel-to-voxel difference metric. Without loss of generality we will report the L1 metric as a measure of the artifact’s severity. L1 is defined as the mean of the absolute voxel-to-voxel difference between the motion simulation and its original version. $L1_{wFT}$, $L1_{wFT2}$ and $L1_{coreg\_shift}$ denoted respectively the artifact severity computed using wFT, wFT2 and coreg\_shift as the reference position.

\section{\textbf{Data}}
\label{data}

\subsection{Synthetic 3D brain data}
\label{synthetic_data}
We relied on synthetic data, since it offers a total control of key image properties: signal to noise ratio (SNR) and contrast. 3D brain data was synthesized with tissue masks extracted from Freesurfer surfaces of gray and white matter (\cite{dahnke2013cortical}). The FSL first tool was used for the estimation of 7 subcortical structures (caudate, putamen, palidum, thalamus, hyppocampy, amygdala, accubens) for each side (left or right). Surfaces were then used by the toblorone software to estimate a Partial volume map for each structure \cite{kirk2019toblerone}. We concatenated the partial volume maps with posterior estimation of cerebellar white and gray matter, skull skin and background extracted by CAT12 segmentation pipeline (\cite{dahnke2013cortical}). 
Each partial volume map was multiplied by a signal value drawn from a Gaussian distribution ($std=1e^{-3}$) and all maps were summed to obtain the synthetic data. This approach was very similar to the one proposed by \cite{billot2021synthseg} to train a segmentation task. The main difference lied in the use of more realistic partial volume maps instead of binary tissue masks. Therefore, no extra smoothing step of the label was required. This provided ideal synthetic acquisition, as if the different tissues were fully homogeneous.
Synthetic data allowed us to study independently different key aspects of the volume: the SNR, the contrast and the specific shape of the brain. The latter can be explored by adding an affine or elastic transformation to the label or by choosing different Human Connectome Projet (HCP) subjects (\cite{glasser2013minimal}). This provided a wide variability of brain shapes. We generated two types of synthetic images using either a T1 like contrast ($CSF=0.1; GM=0.6; WM=1$) or a random contrast where the intensities of each tissue were drawn from a uniform distribution in the $]0, 1[$ range.

\section{\textbf{Results}}
\label{results}

\subsection{1D translation}
\label{1d_translation}

\begin{figure}
  \includegraphics[scale=.40]{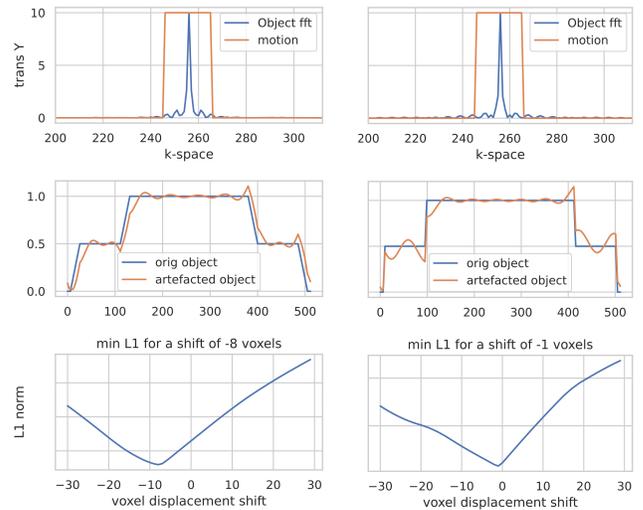}
  \caption{The same motion simulated for two different 1D objects. The first line shows the transient motion step in orange at the k-space center (magnitude of 10 pixels, duration 20), and the FFT of the object in blue (arbitrary units). The middle line represents the original object in blue, and the motion corrupted version in orange. The third line represents the cost function L1 between the original object and its motion-corrupted version as a function of pixel-wise shift. The minimum of this function represents the global displacement induced by the transient motion (coreg\_shift). The two objects (left and right columns) are similar except for the ramp which is of 10 pixels for the left panel and of 2 pixels on the right. }\label{figure1}

\end{figure}

\begin{table}[htp]
\centering
\footnotesize\setlength{\tabcolsep}{2.5pt}

\begin{tabular}{l@{\hspace{7pt}} *{8}{c}}

\toprule
\bfseries  & \textbf{coreg\_shift} &\textbf{wFT\_mean} & \textbf{wFT2\_mean} & \textbf{center} \\
\midrule
\bfseries \textbf{Object with ramp=2} & $1$ & $7.23$ & $9.92$ & $10$ \\ 
\bfseries \textbf{Object with ramp=10} & $8$ & $9.15$ & $9.98$ & $10$  \\
\bottomrule
\end{tabular}
\caption{Displacement in pixels computed for the motion applied on the two objects of Figure \ref{figure1}}\label{table1}
\end{table}

To illustrate the global displacement issue, two different 1D objects were used to simulate a single motion occurring at the center of the k-space. 
\\ The results are shown in Figure \ref{figure1}. The object with a sharp ramp (right column) had a larger spatial frequency content, thus motion in the center of k-space disturbed less of the spatial frequencies of the object (the frequencies outside the motion step in orange are not perturbed, first row). The global shift observed was small (one pixel). For the object with a larger ramp (left column), most of the frequency content of the object was disturbed by motion (orange step, first row), resulting in a global shift of 8 pixels.
\\ The main result was that for a given motion, the induced global displacement varied depending on the original object spatial frequency content. This was counter intuitive because one would have expected that a given movement would move any object by the same amount regardless of its shape. Furthermore, all numerical solutions as defined in section 2.2.2 and reported in Table 1 predicted a large displacement ($> 7 pixels$) for motion occurring in the center of k-space.
\\ With two specific examples, we therefore showed that the theoretical predictions did not correspond to the co-registration ground truth. The 1D case was not further explored as the arbitrary choice of the shape of the object may strongly alter the results. In the next section, we explored the global displacement in the case of 3D motion during the k-space acquisition of a 3D brain MRI.

\subsection{Transient motion during 3D brain acquisition}
\label{transient_3d_motion}

\begin{figure*}
   
  \includegraphics[scale=.89]{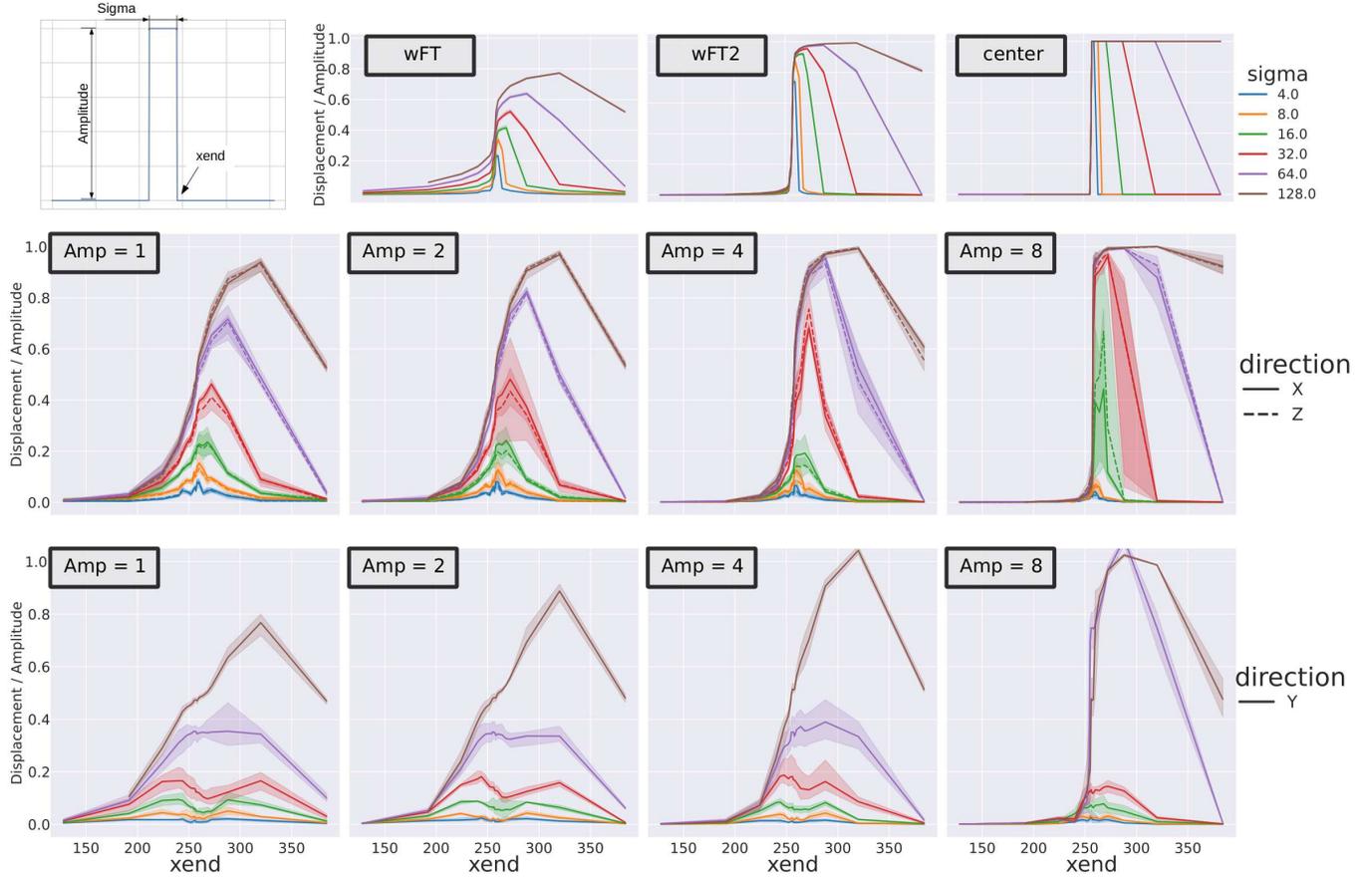}
  \caption{First plot in first row shows the motion time course parametrisation: the displacement amplitude, the duration sigma and the position of k-space xend. The reference position was set as the position at the beginning of the k-space. In the three last plots of first row, the predicted displacements were represented relative to the motion amplitude (in \%) as defined in section \ref{induced_global_disp}. Note that all different amplitude overlap in those cases. In the second and third row we plotted the coreg\_shift value (computed by co-registration relative to input motion amplitude). We compared three different translation directions X, Y, Z. (X and Z on the second row and Y on the third raw). For all simulations the chosen phase encoding direction was the Y-axis. }\label{figure2}

\end{figure*}

In this experiment, we studied the effect of transient motion by considering a two state problem: the subject instantaneously moved from a reference position to a second position for some time and then returned to the original position. The motion time course was defined as a boxcar function. It was thus characterized by four parameters: the displacement direction, the amplitude, the duration (sigma) and the position in the k-space where the subject came back to the initial reference position (xend) as defined in Figure 2. We explored 456 different motion configurations as a combination of 4 amplitudes ([1, 2, 4, 8] mm or degree), 6 durations $sigma = [4, 8, 16, 32, 64, 128]$ and 19 different xend positions from 128 to 384. The duration sigma and the position xend were defined as the number of k-space points for a phase dimension of 512. The motion time course was then interpolated to the real phase dimension of the image (218 in our case). A sigma of 4 and 128 represented a perturbation of 0.8 \% and 25 \% of the total k-space, respectively. We applied each motion to four different synthetic data consisting of simulations of two distinct HCP subjects using two different contrasts (a T1-like contrast and a random contrast).\\

\begin{figure*}
   
    \includegraphics[scale=0.98]{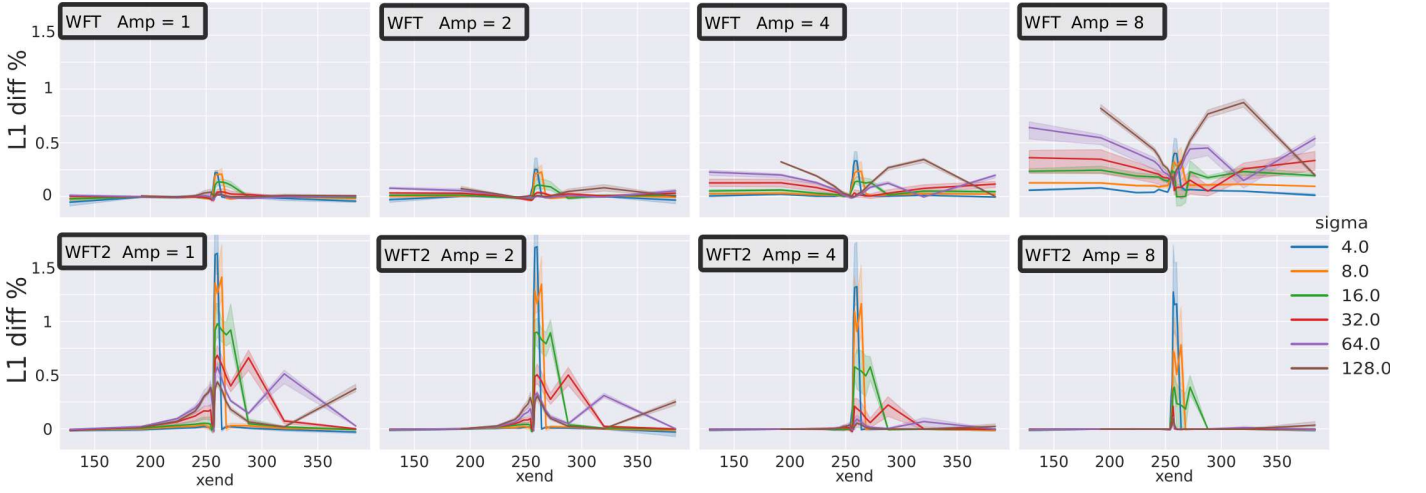}
    \caption{Relative changes of the L1: $\frac{L1_{wFT} – L1_{coreg\_shift}}{L1_{coreg\_shift}}$ (top line) and $\frac{L1_{wFT2} – L1_{coreg\_shift}}{L1_{coreg\_shift}}$ (bottom line). $L1_{wFT}$, $L1_{wFT2}$, $L1_{coreg\_shift}$ are the L1 metrics after simulating motion with different reference positions: wFT, wFT2, coreg\_shift. We simulated the same set of motion as in Figure \ref{figure2}, but only for a translation along the X and Z directions. }\label{figure3}

\end{figure*}

In this experiment we choose always the same reference as the first time point. Prediction of the correct reference is then equivalent to predict the induced global displacement of the motion simulated volume. Figure \ref{figure2} (first row) shows the theoretical displacement as predicted by the different methods (Cf section 2.2.2) for all motion time courses of this experiment. As the displacement was by definition proportional to the amplitude of movement, the predicted shift was represented as a percentage of the input movement amplitude. It then becomes independent of the given amplitude. There was also no dependence on the direction on which the movement occurred, since the averages were computed independently on each Euler parameter. A small dependence on subject shape and contrast was observed, as shown by the small variance over the different simulations (100\% confidence interval), this was due to the weights derived from the Fourier Transform which depended on the subject shape and the image contrast. The predicted displacement by the wFT2 and center methods were very similar and will be considered as equivalent for the rest of this work (we will consider only wFT2).
\\ Figure \ref{figure2} second row, shows the measured coreg\_shift in \% (relative to the input motion amplitude) for the X and Z translations directions, which had similar effects. An important difference when comparing the measured shift to theoretical predictions was the dependence on the amplitude of motion: the 1-mm amplitude looked similar to the wFT predictions, whereas the 8 mm amplitude looked similar to wFT2 predictions. In both cases, there were important differences for short duration motions (sigma < 16) around k-space center: a small shift was measured by the co-registration whereas wFT and wFT2 predicted a large one. 
\\ Figure \ref{figure2} third row, shows the coreg\_shift for a translation in the y direction. The Y translation was very different compared to the X or Z translations, specifically at the k-space center, where almost no shift was measured. This was not due to a spatial specificity of the brain in the y direction, but to the fact that the translation was co-linear to the phase encoding direction. Indeed changing the phase encoding direction for the X axis, then measuring the global shift for the X translation yields similar results to the ones in Figure \ref{figure2} third row (results not shown). This demonstrated a strong interaction between the motion translation direction and the phase encoding direction. To better understand this interaction, we refer the reader to Appendix which shows that applying motion in k-space is equivalent to applying a convolutional filter (Figure \ref{figure6}). 
\\ The results shown in Figure \ref{figure2} demonstrated the discrepancy between the observed global shift (obtained by direct coregistration) and the two theoretical solutions wFT and wFT2. Next, we showed how the choice of a reference influenced the severity of artifacts measured by the L1 difference metric. Motion shown in Figure \ref{figure2} was simulated for three different reference positions (wFT, wFT2 or coreg\_shift) and the L1 metric was computed for each case. Figure \ref{figure3} shows that wFT and wFT2 references lead to higher L1 changes compared to coreg\_shift. We plotted the relative changes $\frac{L1_{wFT} – L1{coreg\_shift}}{L1_{coreg\_shift}}$ for motion translation in the X and Z directions (and the same metrics for wFT2). Positive values were expected as the co-registration algorithm minimized the L1 metric (even though the mutual information was used as cost function) as shown in Figure \ref{figure3}.
\\ We observed larger L1 errors for wFT when applied to large translation amplitudes and for wFT2 when applied to small translation amplitudes. Using wFT or WFT2 as reference lead to an overestimation of the severity of the artifact measured by an L1 metric due to global displacement. 
\\ To summarize, we have shown that wFT or wFT2 predicted a large global displacement for a motion occurring around the k-space center, in contrast to the coregistration that showed no or small displacements for a short transition ($\leq16$). These differences induced an overestimation of the L1 severity. It might have been amplified by the simplicity of the motions simulated. We therefore conducted a last experiment using more realistic random motions.

\subsection{Random motion during 3D brain acquisition}
\label{3d_brain_random_motion}

\begin{figure}
    
    \includegraphics[scale=.35]{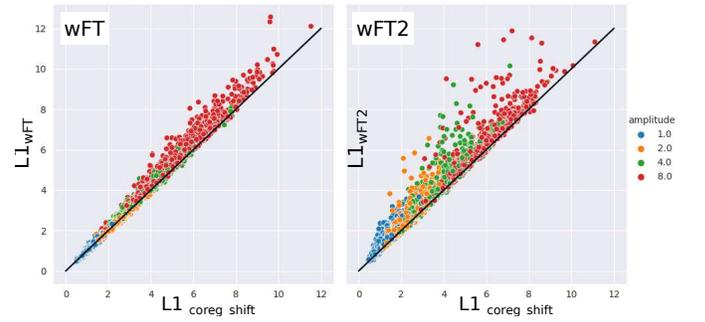}
    \caption{Scatter plots of L1 when choosing wFT as reference on the left and wFT2 on the right, compared with L1 measured by co-registration. 50 random motions of four amplitude levels, on 12 different subjects using 2 types of contrasts were simulated. }\label{figure4}

\end{figure}

\begin{figure}
    
    \includegraphics[scale=.30]{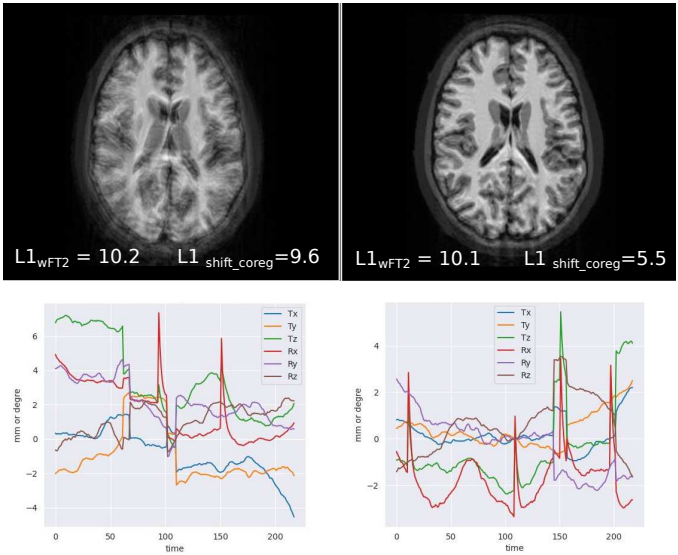}
    \caption{Example of two simulated motion artifacts with the corresponding time courses (amplitude of 8 mm and 8°). Both images have similar L1 when computed with the wFT2 reference. But after co-registration the L1 scores were really different. $L1_{coreg\_shift}$ was more coherent with the visual inspection of both images: the left one looks much more corrupted.  }\label{figure5}

\end{figure}

To further confirm the importance of choosing the correct method for computing the reference position, we proposed to study more realistic motion. We simulated motion as a combination of perlin noise, swallow motions and sudden displacements as explained in section 2.1.2. We compared the L1 values given by the two reference positions: either wFT or wFT2 versus coreg\_shift. As in the previous experiment (Figure \ref{figure3}), we observed larger differences for the wFT2 reference. The wFT method correlated more with the co-registration, but some differences remained: for instance a given motion could lead to $L1_{coreg\_shift}=4$ but $L1_{wFT}=6$ and $L1_{wFT2}=8$. 
\\ To illustrate how the $L1_{wFT2}$ ranking of motion severity can be misleading, we plotted on Figure \ref{figure5}, two motion simulated volumes with the same level of artifact severity ($L1_{wFT2}$ ~ 10) but very different $L1_{coreg\_shift}$ scores. The $L1_{coreg\_shift}$ score was more representative of the visual severity. The discrepancy observed for the volume on the right is due to a mis-coregistration induced by choosing the wFT2 method as a reference. It is worth to note the swallow peak occurring at the k-space center on the randomly simulated time course. This short transient motion around k-space center was the reason why choosing the wFT2 reference (and to a lower extent the wFT reference) induced the global displacement instead of correcting for it. 

\section{\textbf{Discussion}}
\label{discussion}

\begin{figure}[b]
    
    \includegraphics[scale=.65]{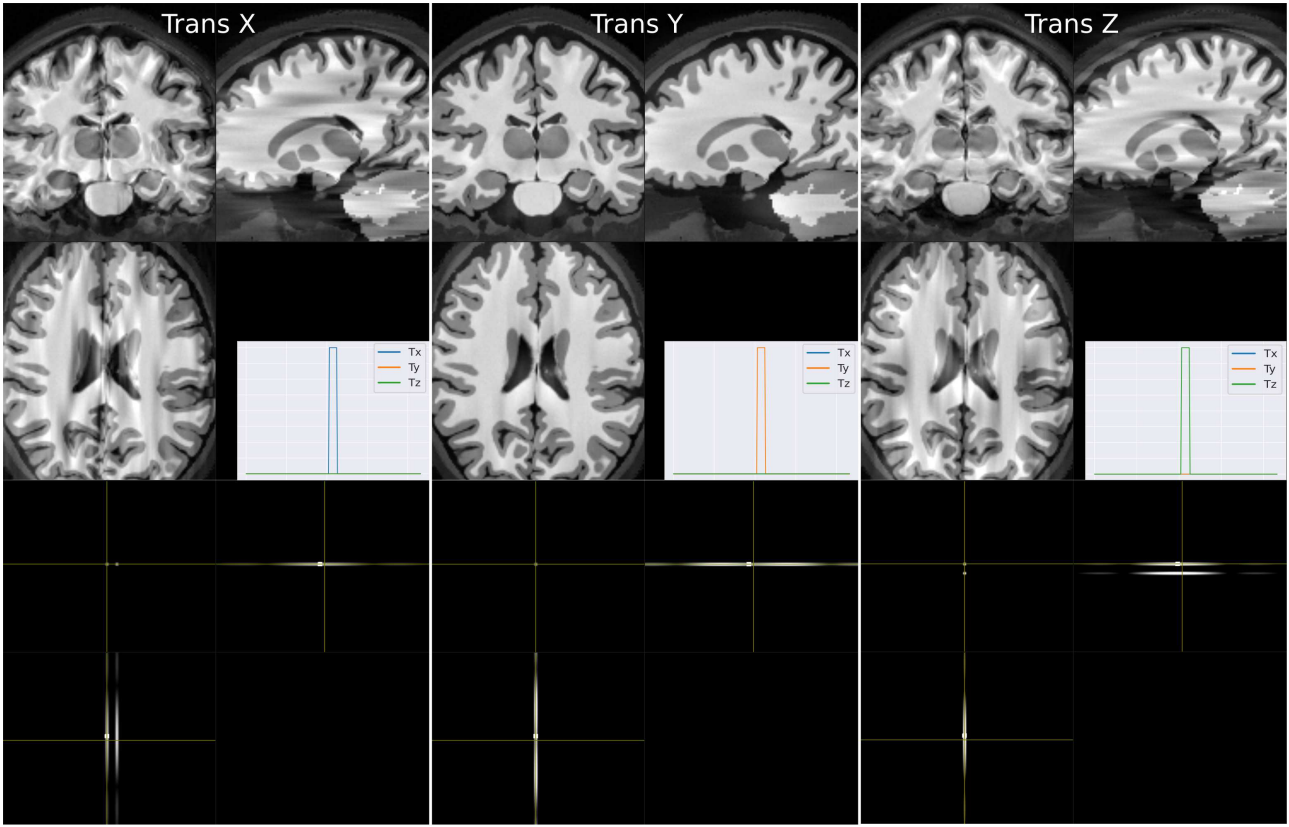}
    \caption{The first row shows the simulated motion corrupted volume for 3 similar input motion time courses: a transient motion at the kspace center ($amplitude=4$, $sigma=10$). Each columns corresponds to a different translation direction, from left to right the X, Y and Z direction. In the second row we plotted the corresponding motion filter G as described in equation \ref{eqfinal}. The slowest phase encoding direction is the Y axis. }\label{figure6}

\end{figure}

All the experiments demonstrated that the global displacement could not always be predicted by the theoretical solutions proposed in the literature (wFT or center). Furthermore using those theoretical references may lead to large overestimation of voxel-to-voxel metrics such as the L1. More specifically we showed that short transient motions around k-space center did not induce a global displacement (only a contrast change). This implied that the reference position estimated with wFT and wFT2 may induce a global shift instead of correcting for it, because their weights are maximum at k-space center. It will then lead to a large overestimation of the L1 metric as demonstrated in Figure \ref{figure3} and \ref{figure5}. in Figure \ref{figure3}, short transitions are for a duration sigma from 4 to 16. For a typical MPRAGE acquisition time of 8 minutes this corresponds to a motion duration of 3.7 to 15 seconds which is long enough to be expected in practice.
\\An even stronger argument is to consider the interaction between the translation direction and the phase encoding direction. When both are aligned, the convolution motion filter that apply to the image does not show two separate lines (Figure \ref{figure6}) compared to the other two translation directions. This is why less spatial mixing and less global displacement are observed in figure \ref{figure2} third row. Since theoretical prediction of global displacement are, by definition, independent of the translation direction, they must be erroneous.
\\Whether or not including an extra co-registration step after the motion simulation will change the deep learning performance is out the scope of this paper, but it will certainly depend on the chosen motion parameters. As we could not derive a proper theoretical or analytical solution we strongly advise to add an extra co-registration step after the simulation when using translation sensitive measures. Note that this is unfortunate, because it increase the computational demand of the motion simulation, and add some noise regard to the exact reference.
\\In this work we considered the Elastix co-registration algorithm as the ground truth to quantify the global displacement. However, as the co-registration results from an optimization method there may be some numerical errors or local minima convergence issues. We visually performed a quality control of different results and did not find any incoherence in our experiments. Even for an amplitude of 1 mm (very small shifts), a visual control of all motions from experience 2 showed an effective global motion which was no more visible after the co-registration. When specific translation directions were studied, the amplitude of the shift in the other directions (which should be zero) were inspected as well and an error of less than 4 \% of the initial amplitude was found. We were then confident that co-registration did not stop in local optimum.
\\In experiment 2, we focused on an over simplified problem: a two position motion. From a theoretical point of view, we expected the reference position to be either position 1 or position 2, depending on how long the object was in each respective position during acquisition. This behavior was more in line with the wFT2 (or center) predictions. But the results for small amplitudes that showed a global displacement in between the two positions were more puzzling. A possible interpretation was that for short amplitudes, the motion artifact induced mainly a local blurring which lead to an in between position in the image domain. However, for larger amplitudes, spatially distinct parts of the object were mixed and the solution was then more binary. With smaller amplitudes, the observed global displacement was more in line with the wFT prediction, except again for short transitions. 
\\The induced global displacement may not be an issue if no difference metric is used. This is indeed the case in the autofocusing litterature, where a global metric was derived from the images and the motion was inferred by minimizing this metric \cite{atkinson1997automatic}. Since no comparison was done with the original image, the global displacement did not matter in this case.

\section{\textbf{Conclusion}}
\label{conclusion}

When dealing with motion artifact simulation, one often needs a measure of the artifact severity which is classically achieved by a voxel-to-voxel difference of the images with and without artifact. In this case it is of particular importance to properly choose the reference position of the motion time course in hand, to avoid any mis-co-registration between the two images. 
\\Theoretical solutions proposed in the literature used the object position around the center of the k-space as a reference: either the position at kpace center (~ wFT2), or a weighted average (wFT). We showed here that this assumption no longer holds for short transient motion at k-space center, where no global displacement was observed. 
\\We also showed a strong dependence of the problem to the amplitude and the direction of the motion which makes a theoretical solution more challenging. It is quite unfortunate not to find any theoretical or analytical solution yet, as this constrains to add a co-registration step which implies an extra computational cost and numerical imprecision.

\newpage


\newpage
\section*{Annex: technical notes : motion simulation in MRI}

\newtheorem{notation}{Notation}
\newtheorem{remark}{Remark}
\def\C{\mathbb C}
\def\R{\mathbb R}
\def\Z{\mathbb Z}
\def\N{\mathbb N}
\def\qed{$\Box$}
\def\phi{\varphi}
\def\epsilon{\varepsilon}
\def\F{{\cal F}}
\def\dd{\textrm d}
\def\ic{\textrm i}
\def\Id{\textrm 1\!\!\!\!1}
\def\sech{\textrm{sech}}

\def\proof{\medskip\noindent{\bf Proof }}
\def\sn{\textrm sn}

\newcommand{\FT}[1]{\ft\left[#1\right]}
\newcommand{\iFT}[1]{\ift\left[#1\right]}
\newcommand{\Exp}[1]{{\rm e}^{#1}}
\newcommand{\bra}[1]{\langle#1|}
\newcommand{\ket}[1]{|#1\rangle}
\newcommand{\obs}[1]{\langle#1\rangle}
\newcommand{\braketA}[3]{\left\langle#1\right|#2\left|#3\right\rangle}
\newcommand{\braketB}[2]{\left\langle#1|#2\right\rangle}
\newcommand{\bec}{\begin{center}}
\newcommand{\enc}{\end{center}}
\newcommand{\be}{\begin{equation}}
\newcommand{\ee}{\end{equation}}
\newcommand{\bmi}{\begin{minipage}}
\newcommand{\emi}{\end{minipage}}
\newcommand{\tw}{\textwidth}
\newcommand{\bi}{\begin{itemize}}
\newcommand{\ei}{\end{itemize}}
\newcommand{\ba}{\begin{array}}
\newcommand{\ea}{\end{array}}
\newcommand{\lt}{\left(}
\newcommand{\rt}{\right)}
\newcommand{\lqu}{\left[}
\newcommand{\rqu}{\right]}
\newcommand{\lgr}{\left\{}
\newcommand{\rgr}{\right\}}
\newcommand{\rd}{\right.}
\newcommand{\bds}{\mathbfsymbol}
\newcommand{\vgrad}[1]{\overrightarrow{\nabla}(#1)}
\newcommand{\bgrad}[1]{\mathbf{\nabla}(#1)}
\newcommand{\Deriv}[2]{\frac{\partial#1}{\partial#2}}
\newcommand{\ora}[1]{\overrightarrow{#1}}
\newcommand{\ud}{\mathrm{d}}
\newcommand{\deriv}[2]{\frac{\ud#1}{\ud#2}}
\renewcommand{\labelitemi}{$\bullet$}
\renewcommand{\labelitemii}{$\ast$}
\renewcommand{\labelitemiii}{$\diamond$}

\everymath{\displaystyle}

\subsubsection*{Numerical formulation of Motion Simulation}
Demonstration of accurate motion simulation start from signal equation : 
\be
	S(\ora{k_0}) = \int_V \ud \ora{r_0} \rho(\ora{r_0})\Exp{\ic \ora{r_0}^t\ora{k_0}}
\ee
Using the following homogeneous coordinate $\mathbf{r}_0=(\ora{r_0},1)=(r_x^0,r_y^0,r_z^0,1)$ and $\mathbf{k}_0=(\ora{k_0},0)=(k_x^0,k_y^0,k_z^0,0)$, the signal equation can be written : 
\be
	S(\mathbf{k}_0) = \int_V \ud \ora{r_0} \rho(\mathbf{r}_0)\Exp{\ic \mathbf{r}_0^t\mathbf{k}_0}
\ee
With homogenous coordinate, motion can be express easily trough use of affine transformation $A$, which express as :
\be
	A(t) = \lt\ba{cc} R(t) & T(t) \\0 & 1\ea\rt,
\ee
with $R(t)$ the rotational compound of motion and $T(t)$ the translational compound and $t$ express the time dependance of motion. The inverse of the affine express as : 
\be
	A^{-1}(t) = \lt\ba{cc} R^{t}(t) & -R^{t}(t)T(t) \\0 & 1\ea\rt
\ee
Effect of motion on the object $\rho$ can be expressed trough action of affine transform as $\rho(A^{-1}(\ora{k_0})\mathbf{r_0})$. The measured signal is thus :
\be
	S(\mathbf{k}_0) = \int_V \ud \ora{r_0} \rho(A^{-1}(\ora{k_0})\mathbf{r}_0)\Exp{\ic\mathbf{r}_0^t\mathbf{k}_0}
\ee
The exponential part is not change as this corresponds to laboratory coordinate, only the object $\rho$ position change : this is the active point of view. Numerical implementation of motion simulation in the active point of view is numerically tedious as the rotated/translated object need to be computed for each k-space points. This amount of computation can be reduced while using the passive point of view. Notes that passive point of view can be expressed in two ways : a first on using NuFFT type 2 (type 2 in the sense from uniform space to a dual non uniform space) from image k-space followed by inverse FFT and a second on that use FFT and then a NuFFT type 1 (type 1 in the sense from non uniform dual space to uniform space) from k-space to image. The first step to express this view point is to use the change of coordiante $\mathbf{r_1}=A^{-1}\mathbf{r_0}$ : 
\begin{align}
	S(\mathbf{k_0}) &= \int_V \ud \ora{r_0} \rho(\mathbf{r}_1)\Exp{\ic (A(\ora{k_0})\mathbf{r}_1)^t\mathbf{k}_0}\\
				&=\int_V \ud \ora{r_0} \rho(\mathbf{r}_1)\Exp{\ic \mathbf{r}_1^tA^t(\ora{k_0})\mathbf{k}_0}\\
				&=\int_V \ud \ora{r_1}\left|J(\ora{r_1})\right| \rho(\mathbf{r}_1)\Exp{\ic \mathbf{r}_1^tA^t(\ora{k_0})\mathbf{k}_0}
\end{align}
The Jacobian $J(\ora{r_1})$ is the identity matrix since $A$ depends only on $\ora{k_0}$ and not on $\ora{r_0}$. The determinant $\left|J(\ora{r_1})\right|$ is one and the signal equation simplify to :
\be
S(\mathbf{k_0})=\int_V \ud \ora{r_1}\rho(\mathbf{r}_1)\Exp{\ic \mathbf{r}_1^tA^t(\ora{k_0})\mathbf{k}_0}
\ee
Expanding the term $ \mathbf{r}_1^tA^t(\ora{k_0})\mathbf{k}_0$ lead to : $\ora{x_1}R^t(\ora{k_0})\ora{k_0}+T^t(\ora{k_0})\ora{k_0}$. The signal equation can now be expressed as : 
\be\label{eqSigShift}
	S(\ora{k_0}) = \Exp{\ic \ora{T}^t(\ora{k_0})\ora{k_0}}\int_V \ud \ora{r_1} \rho(\ora{r_1})\Exp{\ic \ora{r_1}^t\left(R^t\ora{k_0}\right)}
\ee
In the equation above, the term $\left(R^t\ora{k_0}\right)$ corresponds to inhomogeneous grid in discrete case, thus the integral correspond to a non uniform Fourier transform of type 2 (e.g. from uniform to non uniform coordinate). Numerically the implementation is a NuFFT algorithm followed by application of a linear phases. The motion corrupted image is then obtained via an inverse Fourier transform of the signal. Note that a second approach could be derived trough the change of coordinate $\mathbf{k}_1=A^t(\ora{k_0})\mathbf{k}_0$ This will lead to a nufft of type 1 but in this case the Jacobian is no more the identiy matrix and it is difficult to derive an analytic solution.

%
%
%

\section{Effect of translation}
In case of pure translation motion, the effect on the resulting image can be expressed in image domain as a convolution with a motion filter. Starting from equation \ref{eqSigShift}, the assumption of pure translation lead to $R(\ora{r_1})=\Id$ :
\begin{align}
	S(\ora{k_0}) &= \Exp{\ic \ora{T}^t(\ora{k_0})\ora{k_0}}\int_V \ud \ora{r_1} \rho(\ora{r_1})\Exp{\ic \ora{r_1}^t\ora{k_0}}\\
				&= \Exp{\ic \ora{T}^t(\ora{k_0})\ora{k_0}}\FT{\rho(\ora{r_1})}
\end{align}
Doing Fourier transform of this signal and using the convolution properties of Fourier transform we obtain : 
\be\label{eqfinal}
	\tilde{\rho}(\ora{r_1}) = \rho(\ora{r_1})\otimes\iFT{\Exp{\ic \ora{T}^t(\ora{k_0})\ora{k_0}}} = \rho(\ora{r_1})\otimes G.
\ee
Thus in image domain, in case of pure translation, motion acts as a filter $G$ convolving the original images. Plotting this convolutional filter helps to understand why the global displacement is different when the translation direction is the same as the slowest varying phase encoding direction. In the experiment reported in Figure 6, we considered a transient motion occurring around kspace center with an amplitude of 4 and a duration of 10 and we only change the translation direction (X Y and Z). The main observation is that for X and Z direction there is a clear spatial mixing due to the double ray of the filter, whereas only one ray is observed when translation is the same as the phase encoding direction (Y axis)


\end{document}
\endinput